\documentclass[amsmath,amssymb,aps,prd,onecolumn,groupedaddress]{revtex4-2}
\usepackage{bm}
\usepackage{braket}
\usepackage{graphicx}
\usepackage{mathtools}
\usepackage{dcolumn}
\begin{document}
\title{Neutrino mixing angle and neutrino oscillation in ALPs matter}

\author{Alexey Lichkunov}
\email{lichkunov.aa15@physics.msu.ru}
\affiliation{Department of Theoretical Physics, Lomonosov Moscow State University,\\
  119992 Moscow, Russia}
\author{Alexander Studenikin}
\email{studenik@srd.sinp.msu.ru}
\affiliation{Department of Theoretical Physics, Lomonosov Moscow State University,\\
  119992 Moscow, Russia}

\begin{abstract}
Axions and axion-like particles (ALPs) are among of the most popular candidates for dark matter. Axions are also considered as new physics contributions to the muon g – 2 . Following the existed interest to ALPs we consider interaction between neutrinos and hypothetical axion-like particles and derive for the first time the probability of neutrino oscillations accounting for their interactions mediated by ALPs. The corresponding effective mixing angle is derived.
\end{abstract}


\maketitle
\section{Neutrino evolution in ALPs background}
Consider neutrino flavor mixing and oscillations in a background formed by axion-like particles. For simplicity we deal two-flavour neutrino case  ( $\nu_e$ and $\nu_\mu$ ). A similar approach in the case of neutrino propagation in an external electromagnetic field and moving matter can be found in  \cite{Giunti:2014ixa, Pustoshny:2018jxb}.

The Lagrangian of interaction between flavour neutrinos and ALPs can be chosen in the following form (see also \cite{Ma:2022} for the charged lepton interaction with ALPs):
\begin{equation}\label{flag}
L_{int}  = -i\dfrac{\partial_\mu a}{F}\bar \nu_i^{(f)}\gamma^\mu\left({\bar g}_V^{ij} + {\bar g}_A^{ij}\gamma_5\right)\nu_j^{(f)},
\end{equation}
where $\nu_i^{(f)}$ is a neutrino flavor state, $a$ is the axion-like pseudoscalar field, ${\bar g}_V^{ij}$ and ${\bar g}_A^{ij}$ are  the vector and axial-vector interaction constants.

The neutrino flavour states by the mixing matrix $U$ are connected to the neutrino mass states $\nu_i^{(p)}$,
\begin{equation}
\nu^{(f)} = U\nu^{(p)}.
\end{equation}
The Lagrangian (\ref{flag}) can be converted (see, for instance, \cite{Ma:2022}) to:
\begin{equation}
L_{int}  = -i\dfrac{a}{F}\bar \nu_i^{(p)}\left( g_V^{ij}(m_i-m_j) + g_A^{ij}(m_i+m_j)\gamma_5\right)\nu_j^{(p)},
\end{equation}
where $g_V^{ij}$ and $g_A^{ij}$ are the corresponding constants in the basis neutrino mass states.
Using this interaction Lagrangian the following Dirac equation for the neutrino field $\nu_i$ can be derived,
\begin{equation}\label{E}
E_i\nu_i = \left(\gamma_0\mbox{\boldmath$\gamma p_i$}+m_i\gamma_0\right)\nu_i + \dfrac{a}{F}\sum_j\left(g_V^{ij}(m_i - m_j) - g_A^{ij}(m_i+m_j)\gamma_5\right)\gamma_0\nu_j,
\end{equation}
where $E_i$ is the energy of the neutrino mass state.

The shift of the neutrino evolution Hamiltonian in the ALPs background can be calculated by averaging the interaction term in (\ref{E}) (similar to calculations described in detail in \cite{Pustoshny:2018jxb}):
\begin{equation}\label{aver}
\Delta_{ij} = \left\langle\nu_i|H_{int}|\nu_j\right\rangle .
\end{equation}
Finally, for the neutrino mass states evolution Hamiltonian we get
\begin{eqnarray} \label{Hamilt}
    H =
\begin{pmatrix}\label{H}
E_1 & H_{12} \\
H_{21}&E_2
\end{pmatrix},
\end{eqnarray}
where $ H_{ij} = H_{ij}^V + H_{ij}^A $,

\begin{eqnarray}\label{H_V}
H_{ij}^V = \dfrac{a}{F}\left[g_V^{ij}(m_i-m_j)\left(1+\dfrac{p^2}{(E_i+m_i)(E_j+m_j)}\right)
\right]
 \sqrt{\dfrac{(E_i+m_i)(E_j+m_j)}{4E_iE_j}}
\end{eqnarray}
and
\begin{eqnarray}\label{H_A}
H_{ij}^A = -\dfrac{a}{F}\left[g_A^{ij}(m_i+m_j)p\left(\dfrac{1}{E_i+m_i}-\dfrac{1}{E_j+m_j}\right)\right]
 \sqrt{\dfrac{(E_i+m_i)(E_j+m_j)}{4E_iE_j}}.
\end{eqnarray}

\section{Probability of neutrino flavor oscillations}
From (\ref{H}), (\ref{H_V}) and (\ref{H_A}) we get for the effective evolution Hamiltonian for the flavour neutrino states
\begin{equation}
H' = \dfrac{\Delta}{4p}\left(\sigma_1\sin2\theta - \sigma_3\cos2\theta\right) + H_{12}^V\left(\sigma_1\cos2\theta + \sigma_3\sin2\theta\right) - iH_{12}^A\sigma_2,
\end{equation}
where $\theta $ is the neutrino mixing angle in vacuum, $\Delta = m^2_2 -m^2_1$ is the neutrino mass square difference,
and $\sigma_i$ are the Pauli matrices.
For the ultra-relativistic neutrinos ($E \gg m_i$ the Hamiltonian can be approximated by
\begin{eqnarray}
&H_{ij}^V = \dfrac{a}{F}g^{ij}_V(m_i-m_j), \\
&H_{ij}^A = 0.
\end{eqnarray}
The probability of neutrino flavour oscillations in the ALPs background is just straigthforward:
\begin{eqnarray}
&P_{\nu_e \rightarrow \nu_\mu}(x)= \sin^22\theta_{\mathrm{eff}}\sin ^{2} \dfrac{\pi x}{L_{\mathrm{eff}}},
\end{eqnarray}
where the oscillation length is given by
\begin{equation}
L_{\mathrm{eff}} = \dfrac{\pi}{\sqrt{\left(\dfrac{\Delta}{4p}\right)^2 +\left(H_{12}^V\right)^2}}.
\end{equation}
For the effective mixing angle we get
\begin{equation}
\sin^22\theta_{\mathrm{eff}} = \dfrac{\left(\dfrac{\Delta}{4p}\sin2\theta+H_{12}^V\cos2\theta\right)^2}{\left(\dfrac{\Delta}{4p}\right)^2 +\left(H_{12}^V\right)^2}.
\end{equation}
This equation can be rearranged to the form:
\begin{equation}
\sin^22\theta_{\mathrm{eff}} = \sin^22(\theta+\theta_{\mathrm{add}}),
\end{equation}
where
\begin{equation}
\theta_{\mathrm{add}} = \dfrac{1}{2}\arctan\dfrac{4pH_{12}^V}{\Delta} =  \dfrac{1}{2}\arctan\dfrac{4pag^{12}_{V}}{F(m_1+m_2)}.
\end{equation}
It is interesting to note that in the case of the considered neutrino propagation in an ALPs background the neutrino mixing and oscillations exhibit the linear dependence on the neutrino masses.
\normalsize
\vspace{0.5cm}
\section*{Acknowledgments}
The work is supported by the Russian Science Foundation under grant No.22-22-00384.
A.L. acknowledges the support from the National Center for Physics and Mathematics (Project “Study of coherent elastic neutrino-atom and -nucleus scattering and neutrino electromagnetic properties using a high-intensity tritium neutrino source”).

\end{document}